\documentclass{llncs}
\usepackage{graphicx}
\usepackage{cite}
\usepackage{url}
\usepackage{times}
\usepackage{mathptm}

\DeclareSymbolFont{AMSb}{U}{msb}{m}{n}
\DeclareSymbolFontAlphabet{\Bbb}{AMSb}
\def\R{\ensuremath{\Bbb R}}

\let\oldendproof\endproof
\def\endproof{\qed\oldendproof}

\begin{document}

\title{Upright-Quad Drawing of $st$-Planar Learning Spaces} 

\author{David Eppstein}

\institute{Computer Science Department, University of California, Irvine\\
\email{eppstein@uci.edu}}

\maketitle   

\begin{abstract}
We consider graph drawing algorithms for learning spaces, a type of $st$-oriented partial cube derived from antimatroids and used to model states of knowledge of students. We show how to draw any $st$-planar learning space so all internal faces are convex quadrilaterals with the bottom side horizontal and the left side vertical, with one minimal and one maximal vertex. Conversely, every such drawing represents an $st$-planar learning space.
We also describe connections between these graphs and arrangements of translates of a quadrant.
\end{abstract}

\section{Introduction}

A \emph{partial cube} is a graph that can be given the geometric structure of a hypercube, by assigning the vertices bitvector labels in such a way that the graph distance between any pair of vertices equals the Hamming distance of their labels. Partial cubes can be used to describe benzenoid systems in chemistry \cite{ImrKla-00}, weak or partial orderings modeling voter preferences in multi-candidate elections \cite{HsuFalReg-02}, integer partitions in number theory \cite{Epp-06-ip}, and the hyperplane arrangements familiar to computational geometers \cite{Ovc-DAM-06,math.CO/0510263}.
In previous work we found algorithms for drawing arbitrary partial cubes, as well as partial cubes that have drawings as planar graphs with symmetric faces~\cite{Epp-GD-04}.

Here we consider graph drawing algorithms for \emph{learning spaces} (also called \emph{knowledge spaces}), a type of partial cube derived used to model states of knowledge of students~\cite{DoiFal-99}. These graphs can be large; Doignon and Falmagne~\cite{DoiFal-99} write ``the number of knowledge states obtained for a domain containing 50 questions in high school mathematics ranged from about 900 to a few thousand.'' Thus, it is important to have efficient drawing techniques that can take advantage of the special properties of these graphs.

Our goal in graph drawing algorithms for special graph families is to combine the standard graph drawing aesthetic criteria of vertex separation, area, etc., with a drawing style from which the specific graph structure we are interested in is visible. Ideally, the drawing should be of a type that exists only for the graph family we are concerned with, so that membership in that family may be verified by visual inspection of the drawing. For instance, in our previous work~\cite{Epp-GD-04}, the existence of a planar drawing in which all faces are symmetric implies that the graph of the drawing is a partial cube, although not all partial cubes have such drawings. Another result of this type is our proof~\cite{EppGooMen-GD-05} that the graphs having delta-confluent drawings are exactly the distance-hereditary graphs.

The learning spaces considered in this paper are directed acyclic graphs with a single source and a single sink. It is natural, then, to consider $st$-planar learning spaces, those for which there exists a planar embedding with the source and sink on the same face. As we show, such graphs can be characterized by drawings of a very specific type: Every $st$-planar learning space has a dominance drawing in which all internal faces are convex quadrilaterals with the bottom side horizontal and the left side vertical. 
We call such a drawing an {\em upright-quad drawing}, and we describe linear time algorithms for finding an upright-quad drawing of any $st$-planar learning space. Conversely, every upright-quad drawing comes from an $st$-planar learning space in this way.

\section{Learning Spaces}

Doignon and Falmagne~\cite{DoiFal-99} consider sets of concepts that a student of an academic discipline might learn, and define a {\em learning space} to be a family $\cal F$ of sets modeling the possible states of knowledge that a student could have.  Some concepts may be learnable only after certain prerequisites have been learned, so $\cal F$ may not be a power set. However, there may be more than one way of learning a concept, and therefore more than one set of prerequisites the knowledge of which allows a concept to be learned. We formalize these intuitive concepts mathematically with the following axioms:

\begin{description}
\item[\textbf{[L1]}]
If $S\in\cal F$ and $S\neq\emptyset$, then there exists $x\in S$ such that $S\setminus\{x\}\in\cal F$. That is, any state of knowledge can be reached by learning one concept at a time.
\item[\textbf{[L2]}]
If $S$, $S\cup\{x\}$, and $S\cup\{y\}$ belong to $\cal F$, then $S\cup\{x,y\}\in\cal F$.
That is, learning one concept cannot interfere with the ability to learn a different concept.
\end{description}

These axioms characterize families $\cal F$ that form {\em antimatroids}~\cite[Lemma III.1.2]{KorLovSch-91}.
We define a \emph{learning space} to be a graph having one vertex for each set in an antimatroid $\cal F$, and with a directed edge from each set $S\in\cal F$ to each set $S\cup\{x\}\in\cal F$. If $U=\bigcup\cal F$, we say that it is a learning space \emph{over $U$}.
Antimatroids also arise in other contexts than learning; e.g., the family of intersections of a point set in $\R^d$ with complements of convex bodies forms an antimatroid.
In the remainder of this section we outline some standard antimatroid theory needed for the rest of our results.

\begin{lemma}
\label{lem:K1}
If $\cal F$ satisfies axioms L1 and L2, and $K \subset L$ are two sets in $\cal F$, with $|L\setminus K| = n$, then there is a chain of sets 
$K_0=K \subset  K_1\subset\cdots\subset K_n = L$, all belonging to $\cal F$,
such that $K_{i} = K_{i-1} \cup \{q_i\}$ for some $q_i$.
\end{lemma}

\begin{proof}
We use induction on $|K|+|L|$.
If $K$ is empty, let $x$ be given by axiom L1 for $S=L$, and combine $q_n=x$ with the chain formed by induction for $K$ and $L\setminus\{x\}$. Otherwise, let $x$ be as given by axiom L1 for $S=K$, and form by induction a chain from $K\setminus\{x\}$ to $L$.
By repeatedly applying axiom L2 we may add $x$ to each member of this chain not already containing it, forming a chain
with one fewer step from $K$ to $L$.
\end{proof}

Similar repetitive applications of axiom~L2 to the chain resulting from Lemma~\ref{lem:K1} proves the following:

\begin{lemma}
\label{lem:K2}
If $\cal F$ satisfies axioms L1 and L2, and $K \subset L$ are two sets in $\cal F$, with $K \cup \{q\} \in\cal F$ and $q\notin L$, then  $L \cup \{q\} \in\cal F$.
\end{lemma}

\begin{lemma}[Cosyn and Usun~\cite{CosUzu-05}]
\label{lem:closed-wg}
Let $\cal F$ satisfy the conclusions of Lemmas~\ref{lem:K1} and~\ref{lem:K2}. Then the union of any two members of $\cal F$ also belongs to $\cal F$, and $\cal F$ is \emph{well-graded}; that is, that any two sets in $\cal F$ can be connected by a sequence of sets, such that any two consecutive sets in the sequence differ by a single element and the length of the sequence equals the size of the symmetric difference of the two sets.
\end{lemma}

As their edges are oriented from smaller sets to larger ones, learning spaces are directed acyclic graphs.  Lemma~\ref{lem:closed-wg} implies that any learning space~$G$ is a partial cube when viewed as an undirected graph.  Axiom~1 implies that the empty set belongs to $\cal F$, and that any other set has an incoming edge; that is, the empty set forms the unique source in $G$.  Closure under unions implies that $\bigcup\cal F$ is the unique sink in $G$.
Thus, $G$ is \emph{$st$-oriented} (or has a \emph{bipolar orientation}): it is a DAG with a single source and a single sink.  In this paper we are particularly concerned with learning spaces for which this orientation is compatible with a planar drawing of the graph, in that the source and sink can both be placed on the outer face of a planar drawing. A graph admitting such a drawing is an \emph{$st$-planar learning space}.

\section{Upright-Quad Drawings}

In any point set in the plane, we say that $(x,y)$ is \emph{minimal} if no point $(x',y')$ in the set has $x'<x$ or $y'<y$, and \emph{maximal} if no point $(x',y')$ in the set has $x'>x$ or $y'>y$.

\begin{figure}[t]
\centering
\includegraphics[width=1.5in]{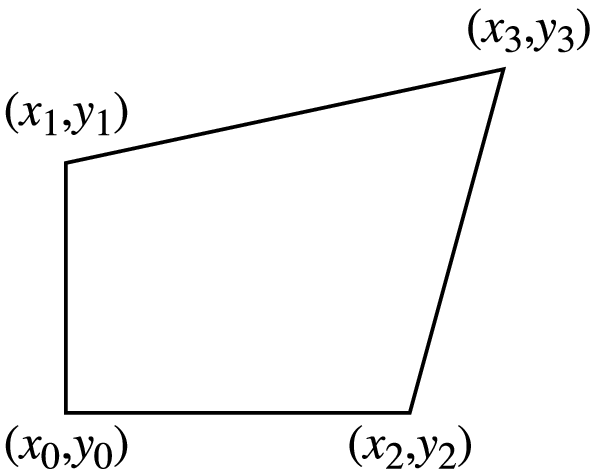}
\qquad\qquad
\includegraphics[width=1.5in]{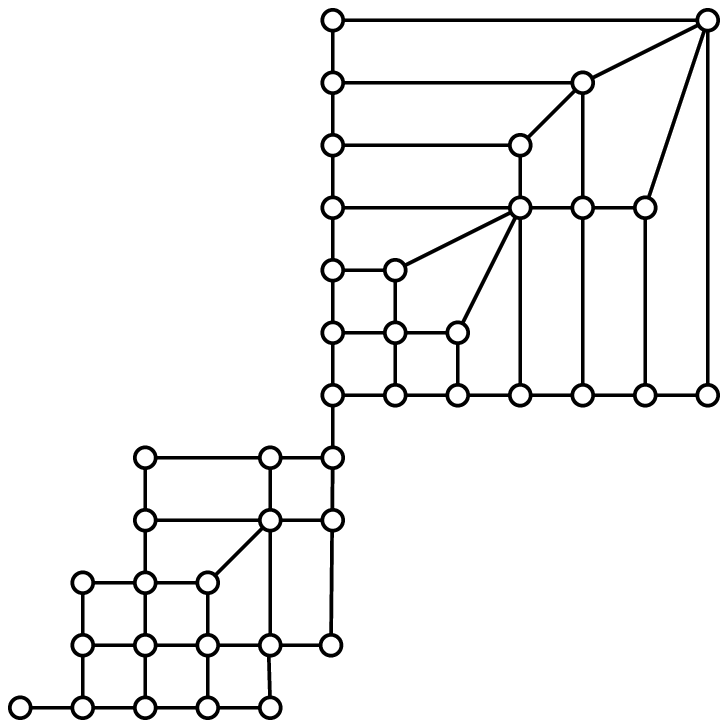}
\caption{Left: An upright quadrilateral. Right: An upright-quad drawing.}
\label{fig:urquad-uqd}
\end{figure}

We define an \emph{upright quadrilateral} to be a convex quadrilateral with a unique minimal vertex and a unique maximal vertex, such that the edges incident to the minimal vertex are horizontal and vertical. That is, it is the convex hull of four vertices $\{(x_i,y_i)\mid 0\le i<4\}$ where $x_0=x_1<x_2\le x_3$ and $y_0=y_2<y_1\le y_3$ (Figure~\ref{fig:urquad-uqd}(left)). We define the \emph{bottom edge} of an upright quadrilateral to be the horizontal edge incident to the minimal vertex, the \emph{left edge} to be the vertical edge incident to the minimal vertex, and the \emph{top edge} and \emph{right edge} to be the edges opposite the bottom and left edges respectively.

We define an {\em upright-quad drawing} of a graph $G$ to be a placement of the vertices of the graph in the plane, with the following properties:

\begin{description}
\item[\textbf{[U1]}]
The placement forms a planar straight line drawing. That is, any two vertices are assigned distinct coordinates, and if the edges of $G$ are drawn as straight line segments then no two edges intersect except at their endpoints.

\item[\textbf{[U2]}]
There is a unique vertex of $G$ that is the minimal point among the locations of its neighbors in $G$, and a unique vertex of $G$ that is the maximal point among the locations of its neighbors in $G$.

\item[\textbf{[U3]}]
Every interior face of the drawing is an upright quadrilateral, the sides of which are edges of the drawing.\end{description}

In an upright-quad drawing, all edges connect a pair of points $(x,y)$ and $(x',y')$ with $x'\le x$ and $y'\le y$; if we orient each such edge from $(x',y')$ to $(x,y)$ then the resulting graph is directed acyclic with a unique source and sink.
As we now show, with this orientation the drawing is a {\em dominance drawing}: that is, the dominance relation in the plane and the reachability relation in the graph coincide.

\begin{lemma}
\label{lem:uq-is-dom}
For any two vertices $(x',y')$ and $(x,y)$ in an upright-quad drawing, $(x'\le x)\wedge (y'\le y)$ if and only if there exists a directed path in the orientation specified above from $(x',y')$ to $(x,y)$.
\end{lemma}

\begin{proof}
In one direction, if there exists a directed path from $(x',y')$ to $(x,y)$, then each edge in the path steps from a vertex to another vertex that dominates it, and the result holds by transitivity of dominance.

In the other direction, suppose that $(x,y)$ dominates $(x',y')$; we must show the existence of a directed path from $(x',y')$ to $(x,y)$. To do so, we show that we can find an outgoing edge to another vertex dominated by $(x,y)$; the result follows by induction on the number of vertices. First consider the case that $(x',y')$ is the minimal corner of some upright quadrilateral of the drawing; then there exist both horizontal and vertical outgoing edges from $(x',y')$. $(x,y)$ cannot belong to the bounding rectangle of these two edges, for if it did we could not use them as part of an empty upright quadrilateral, so at least one of the two edges leads to another vertex that is also dominated by $(x,y)$.

In the second case, suppose $(x',y')$ belongs to the bottom side of some upright quadrilateral of the drawing but is not the minimal vertex of that face. The sequence of faces of the drawing on a vertical line through $x'$, above $(x',y')$, must project to a sequence of nested intervals on the $y$ axis, for each consecutive pair of faces shares an edge which has the same projection onto the $y$ axis as the lower of the two faces. Therefore, $(x,y)$ cannot project to a point interior to the projection of the face above $(x',y')$, so the horizontal outgoing edge from $(x',y')$ leads to a vertex that is also dominated by $(x,y)$.
The case that $(x',y')$ belongs to the left side of an upright quadrilateral but is not its minimal vertex is symmetric to this one.

Finally, suppose that $(x',y')$ is not on the bottom or left side of any upright quadrilateral. Then there can only be a single edge outgoing from $(x',y')$, and there can be no interior faces of the drawing directly above or to the right of this edge. Thus, again, $(x,y)$ cannot project into the interior of the projection of this edge in either coordinate axis, so this edge leads to a vertex that is also dominated by $(x,y)$.
\end{proof}

As is well known to the graph drawing community~\cite{DiBEadTam-99,DiBTamTol-DCG-92}, a dominance drawing exists for any $st$-oriented plane graph in which the source~$s$ and sink~$t$ of the orientation belong to the outer face of the plane embedding; such a graph is known as an \emph{$st$-planar graph}. However, due to property U3, not every $st$-planar dominance drawing is an upright-quad drawing.

\section{Arrangements of Quadrants}

Consider a collection of convex wedges in the plane, all translates of each other.  Recall~\cite{Epp-GD-04,FraOss-GD-03} that a {\em weak pseudoline arrangement} is a collection of curves in the plane, each topologically equivalent to a line and extending to infinity at both ends, such that any two non-disjoint curves meet in a single crossing point.   If no two wedges have boundaries on the same line, the boundary curves of the wedges form such an arrangement, for any two translates of the same wedge can only meet in a single crossing point.

\begin{figure}[t]
\centering\includegraphics[width=3in]{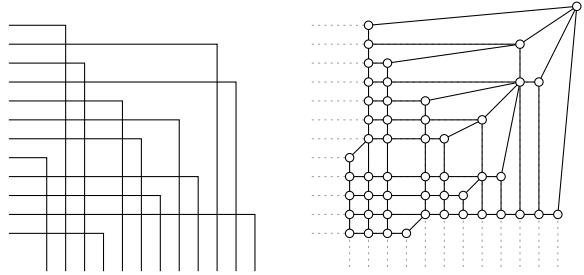}
\caption{Left: an arrangement of quadrants. Right: the region graph of the arrangement, drawn with each vertex (except the top right one) at the maximal point of its region.}
\label{fig:qa2uqd}
\end{figure}

By an appropriate linear transformation, we may transform our wedges to any desired orientation and convex angle, without changing the combinatorics of their arrangement. For later convenience, we choose a standard form for such arrangements in which each wedge is a translate of the negative quadrant $\{(x,y)\mid x,y\le 0\}$ (Figure~\ref{fig:qa2uqd}(left)). For such wedges, the condition that no two quadrants share a boundary line is equivalent to all translation vectors having distinct $x$ and $y$ coordinates. We call an arrangement of translated negative quadrants satisfying this distinctness condition an \emph{arrangement of quadrants}. We refer to the curves of the arrangement, and to the wedges they form the boundaries of, interchangeably.

As with any arrangement of curves, we may define a \emph{region graph} that is the planar dual of the arrangement: it has one vertex per region of the arrangement, with two vertices adjacent whenever the corresponding regions are adjacent across a nonzero length of curve of the arrangement.
For our arrangements of quadrants, it is convenient to draw the region graph with each region's vertex in the unique maximal point of its region, except for the upper right region which has no maximal point.
We draw the vertex for the upper region at any point with $x$ and $y$ coordinates strictly larger than those of any curve in our arrangement. The resulting drawing is shown in Figure~\ref{fig:qa2uqd}(right).

\begin{theorem}
\label{thm:qa2uqd}
The placement of vertices above produces an upright-quad drawing for the region graph of any arrangement of quadrants.
\end{theorem}

\begin{proof}
The drawing's edges consist of all finite segments of the arrangement curves, together with diagonal segments connecting corners of arrangement curves within a region to the region's maximal point; therefore it is planar. Each finite region of the arrangement is bounded above and to the right by a quadrant, either a single curve of the arrangement or the boundary of the intersection of two of the wedges of the arrangement. Each finite region is also bounded below and to the left by a staircase formed by a union of wedges of the arrangement; the drawing's edges subdivide this region into upright quadrilaterals by diagonals connecting the concave corners of the region to its maximal point. A similar sequence of upright quadrilaterals connects the staircase formed by the union of all arrangement wedges to the point representing the upper right region, which is the unique maximal vertex of the drawing. The unique minimal vertex of the drawing represents the region formed by the intersection of all arrangement wedges. Thus, all requirements of an upright-quad drawing are met.
\end{proof}

\begin{theorem}
\label{thm:qa2stpls}
The region graph of any arrangement of quadrants can be oriented to represent an $st$-planar learning space.
\end{theorem}

\begin{proof}
We associate with each vertex of the region graph the set of wedges that do not contain any point of the region corresponding to the vertex. Each region other than the one formed by intersecting all wedges of the arrangement (associated with the empty set) has at least one arrangement curve on its lower left boundary; crossing that boundary leads to an adjacent region associated with a set of wedges omitting the one whose boundary was crossed; therefore axiom L1 of a learning space is satisfied.

If a region of the arrangement, associated with set $S$, has a single arrangement curve $c$ as its upper right boundary, then all supersets of $S$ associated with other regions are also supersets of $S\cup\{c\}$. Thus, in this case, there can be no two distinct sets $S\cup\{x\}$ and $S\cup\{y\}$ in the family of sets associated with the region graph, and axiom L2 of a learning space is satisfied vacuously.

On the other hand, if a region $r$ of the arrangement, associated with set $S$, has curve $x$ as its upper boundary and curve $y$ as its right boundary, then the only sets in the family formed by adding a single element to $S$ can be $S\cup\{x\}$ and $S\cup\{y\}$. In this case,
the region diagonally opposite $r$ across the vertex where $x$ and $y$ meet is associated with the set $S\cup\{x,y\}$ and again axiom L2 of a learning space is met.
\end{proof}

Not all upright-quad drawings are formed from arrangements of quadrants as described here. For instance, a single square is itself an upright-quad drawing, but not one formed in this way. Nevertheless, as we describe in the rest of the paper, Theorems \ref{thm:qa2uqd} and \ref{thm:qa2stpls} have converses, in that any $st$-planar learning space can be given an upright-quad drawing and any upright-quad drawing is combinatorially equivalent to the region graph of an arrangement of quadrants. Thus, these three seemingly different concepts, $st$-planar learning spaces, upright-quad drawings, and region graphs of arrangements of quadrants, are shown to be three faces of the same underlying mathematical objects.

\section{Drawing $st$-Planar Learning Spaces}

\begin{figure}[t]
\centering\includegraphics[width=3in]{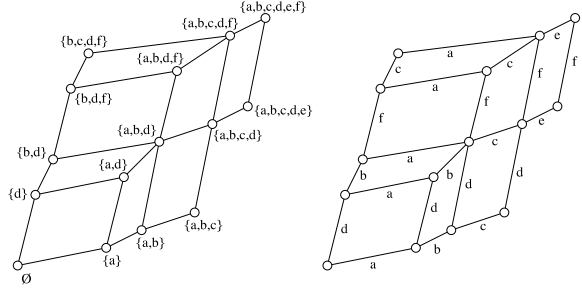}
\caption{An $st$-planar learning space.}
\label{fig:stpls}
\end{figure}

\begin{figure}[t]
\centering\includegraphics[width=1.25in]{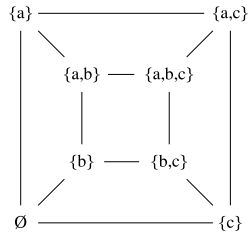}
\caption{A learning space that is planar but not $st$-planar (the power set on three elements).}
\label{fig:nonstp}
\end{figure}

As we have seen, learning spaces are $st$-oriented. Thus, when considering drawing algorithms for these graphs, it is natural to consider the special case in which the $st$-orientation is consistent with a planar embedding; that is, when the graph is $st$-planar. As we show in this section, every $st$-planar learning space has an upright-quad drawing. An example of an $st$-planar learning space is shown in Figure~\ref{fig:stpls}; in the left view, the vertices of a dominance drawing of the graph are labeled by the corresponding sets in the family $\cal F$, while on the right view, each edge is labeled by the single element by which the sets at the two ends of the edge differ. However, not all planar learning spaces are $st$-planar; Figure~\ref{fig:nonstp} shows an example of a learning space that is planar but not $st$-planar.

\begin{lemma}
\label{lem:stpls-quad-face}
Let $G$ be an $st$-planar learning space. Then every interior face of $G$ is a quadrilateral,
with equal labels on opposite pairs of edges.
\end{lemma}

\begin{proof}
Let $b$ be the bottom vertex of any interior face~$f$, with outgoing edges to $b\cup\{x\}$ and $b\cup\{y\}$.
Then by axiom L2 of learning spaces, $G$ must contain a vertex $b\cup\{x,y\}$.
This vertex must be the top vertex of $f$, for otherwise the edges from $b\cup\{x\}$ to $b\cup\{x,y\}$ and from $b\cup\{y\}$ to $b\cup\{x,y\}$ would have to pass above the top vertex, implying a subset relationship from the top vertex to $b\cup\{x,y\}$, which is absurd.
\end{proof}

\begin{figure}[t]
\centering\includegraphics[width=1.5in]{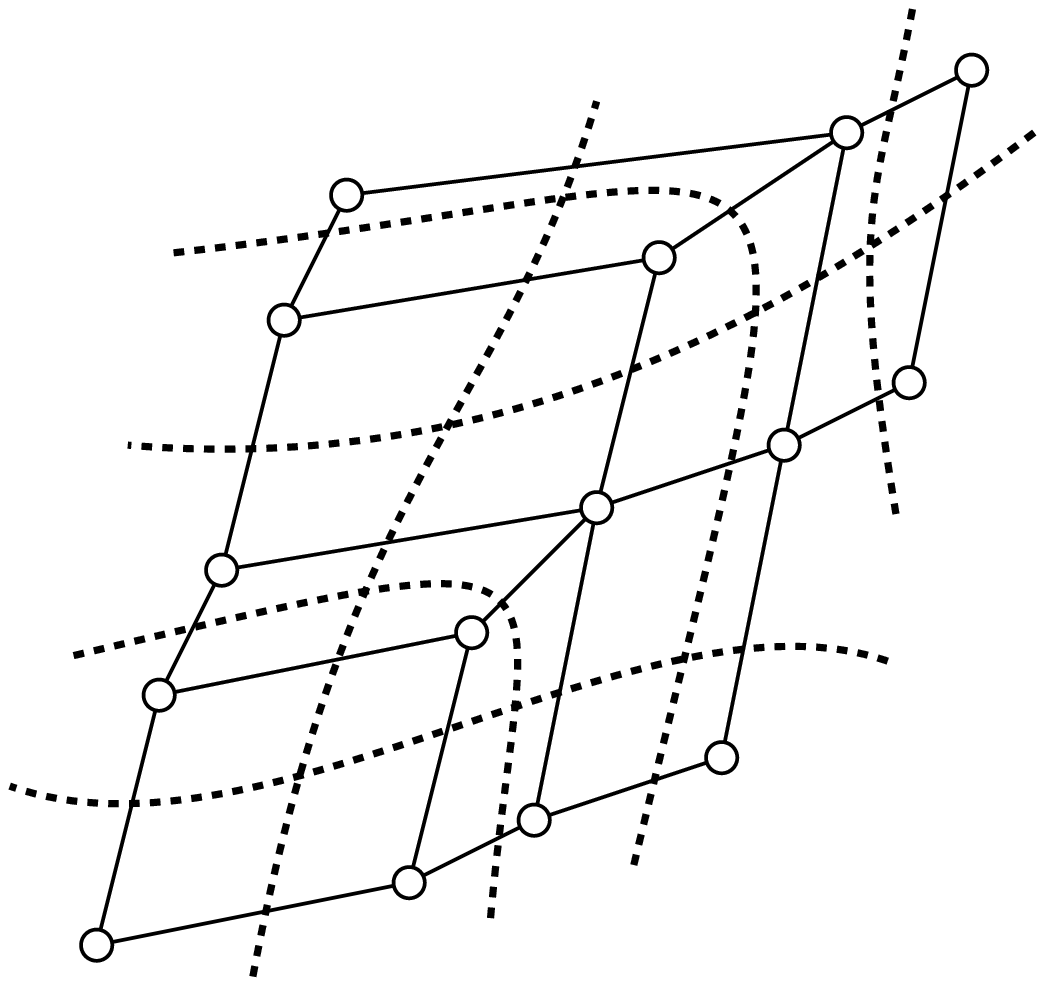}
\qquad\includegraphics[width=1.25in]{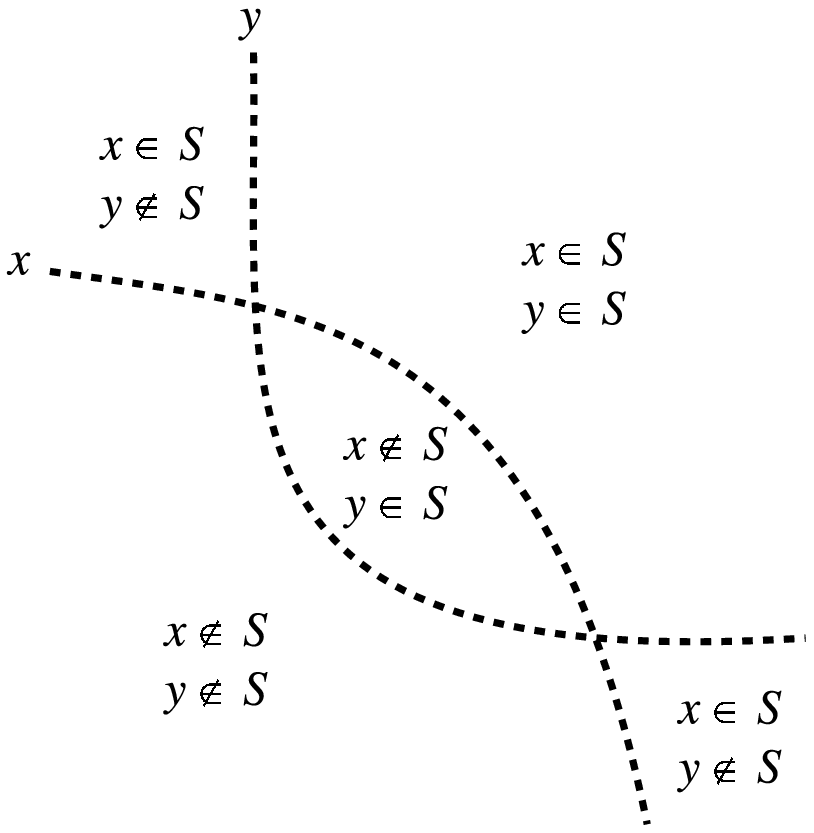}
\caption{Left:The curve arrangement ${\cal A}(G)$ dual to an $st$-planar learning space.
Right: Two crossings between the same two curves lead to a contradiction, so $\cal A$ must be a weak pseudoline arrangement (Lemma~\ref{lem:stpls2wpla}).}
\label{fig:stpls2wpla}
\end{figure}

Define a {\em zone} of a label $x$ in an $st$-planar learning space $G$ to be the set of interior faces containing edges labeled by~$x$. By Lemma~\ref{lem:stpls-quad-face}, zones consist of chains of faces linked by opposite pairs of edges. We may form a curve arrangement ${\cal A}(G)$ from an $st$-planar graph $G$ by drawing a curve through each face of each zone, crossing only edges of $G$ with the label of the zone. Within each face, there are two curves, which we may draw in such a way that they cross once; they may also be extended to infinity past the exterior edges of the drawing without any crossings in the exterior face (Figure~\ref{fig:stpls2wpla}(left)).
${\cal A}(G)$ can be viewed as a form of planar dual to $G$, in that it has one vertex within each face of $G$, one face containing each vertex of $G$, and one arrangement segment crossing each edge of $G$; however it lacks a vertex dual to the outer face of $G$.

\begin{lemma}
\label{lem:stpls2wpla}
If $G$ is an $st$-planar learning space, then ${\cal A}(G)$ is a weak pseudoline arrangement.
\end{lemma}

\begin{proof}
The curves in  ${\cal A}(G)$ are topologically equivalent to lines and meet only at crossings.
Suppose for a contradiction that two curves labeled $x$ and $y$ in ${\cal A}(G)$ cross more than once.
Then (Figure~\ref{fig:stpls2wpla}(right))
two different regions between these curves would contain vertices corresponding to sets containing $x$ and not containing $y$ or vice versa. But then every path from one such set to another would cross one or the other of the two curves, contradicting the assumption that $G$ is a partial cube and has shortest paths labeled only by the elements of the symmetric difference of the two path endpoints.
\end{proof}

\begin{figure}[t]
\centering\includegraphics[width=3.5in]{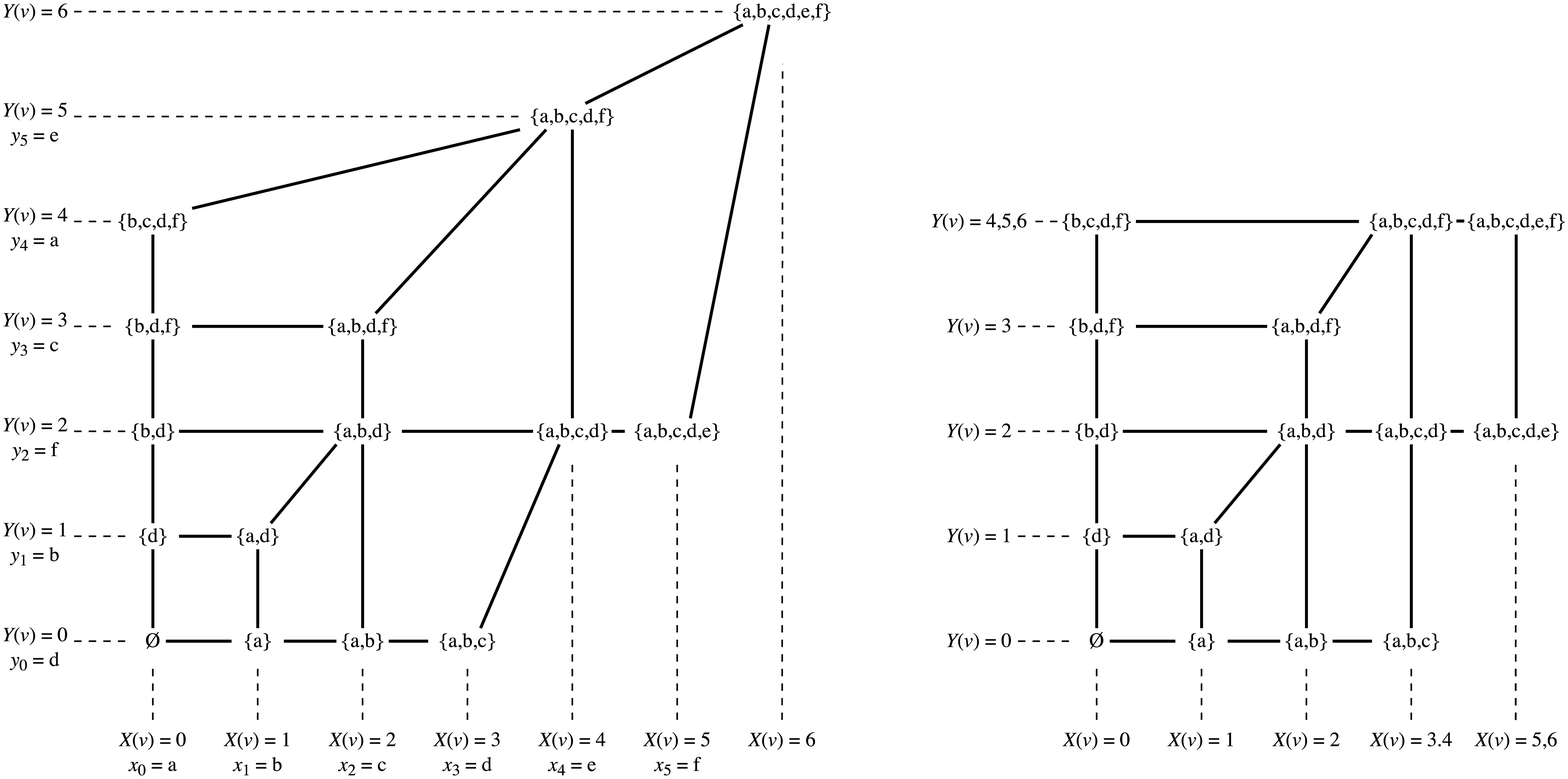}
\caption{Left: coordinates for conversion of $st$-planar learning space to upright-quad drawing. Right: the same drawing with compacted coordinates.}
\label{fig:stpls2uq}
\end{figure}

We are now ready to define the vertex coordinates for our upright-quad drawing algorithm.
Consider the sequence of labels $x_0, x_1,\ldots x_{\ell-1}$ occurring on the right path from the bottom to the top vertex of the external face of the drawing.
For any vertex $v$ of our given $st$-planar learning space, let $X(v)=\min\{i\mid x_i\notin v\}$. If $v$ is the topmost vertex of the drawing, define instead $X(v)=\ell$.
Similarly, consider the sequence of labels $y_0, y_1,\ldots y_{\ell-1}$ occurring on the left path from the bottom to the top vertex of the external face of the drawing.
For any vertex $v$ of our given $st$-planar learning space, let $Y(v)=\min\{i\mid y_i\notin v\}$. If $v$ is the topmost vertex of the drawing, define instead $Y(v)=\ell$.

\begin{lemma}
\label{lem:crossing-order}
Let $G$ be an $st$-planar learning space, with $y_i$ as above,
let $i<j<k$, and suppose that the curves labeled $y_i$ and $y_k$ both cross the curve labeled $y_j$ in the arrangement ${\cal A}(G)$. Then the crossing with $y_k$ occurs to the left of the crossing with $y_i$.
\end{lemma}

\begin{proof}
Otherwise, the arrangement would contain a set containing $y_i$ but not $y_j$ in the region left of the crossing between $y_i$ and $y_j$, and a set containing $y_k$ but not $y_j$ in the region right of the crossing between $y_j$ and $y_k$. However, as $y_i$ and $y_k$ could only cross above $y_k$, there could be no sets containing both $y_i$ and $y_k$ but not $y_j$, violating the closure of a learning space's sets under union (Lemma~\ref{lem:closed-wg}).
\end{proof}

\begin{lemma}
\label{lem:stpls2uq}
If we place each vertex $v$ of an $st$-planar learning space $G$ at the coordinates $(X(v),Y(v))$, the result is an upright-quad drawing of $G$.
\end{lemma}

\begin{proof}
It is clear from the definitions that each edge of $G$ connects vertices with monotonically nondecreasing coordinates.  We show that each internal face is an upright quadrilateral. Consider any such face $f$, with bottom vertex $b$, top vertex $t$, bottom and top edges labeled $x_i$, and left and right edges labeled $y_j$. Then, for any edge label $y_k$ with $k<j$, $y_k\in b$;
for otherwise, the curve for $y_k$ would cross the curve for $y_j$ to the right of $f$, and curves $x_i$, $y_j$, and $y_k$ would violate Lemma~\ref{lem:crossing-order}. Thus, the vertices $b$ and $b\cup\{x\}$ of $f$ are placed at $y$-coordinate value $j$, and the other two vertices have $y$-coordinates larger than $j$.
Symmetrically, the vertices $b$ and $b\cup\{y\}$ of $f$ have $x$-coordinate $i$, and the other two vertices have $x$-coordinates larger than~$i$.

This shows that all edges that are bottom or left edges of an interior face of the drawing are horizontal or vertical. If $e$ is not such an edge, then it belongs to the left or right exterior path of the drawing. If on the left path, it connects a vertex $\{y_i' \mid i'<i\}$ to $\{y_i' \mid i'\le ii\}$ and thus has strictly increasing $y$ coordinates; symmetrically, if on the right path, it has strictly increasing $y$ coordinates. Thus all such edges also have the correct dominance order for their vertices.

As all edges are oriented correctly, the drawing must have a unique minimal vertex and a unique maximal vertex, the source and sink of $G$ respectively. Together with each face being an upright quadrilateral, this property shows that the drawing is an upright-quad drawing.
\end{proof}

A drawing produced by the technique of Lemma~\ref{lem:stpls2uq} is shown in Figure~\ref{fig:stpls2uq}(left). As in standard $st$-planar dominance drawing algorithms~\cite{DiBEadTam-99}, we may compact the drawing by merging coordinate values $X(v)=i$ and $X(v)=i+1$ whenever the merge would preserve the dominance ordering of the vertices; a compacted version of the same drawing is shown on the right of Figure~\ref{fig:stpls2uq}.

\begin{theorem}
\label{thm:stpls2uq}
Every $st$-planar learning space $G$ over a set $U$, having $n$ vertices, has an upright-quad drawing in an integer grid of area $(|U|+1)^2$ that may be found in time $O(n)$.
\end{theorem}

\begin{proof}
We construct an $st$-planar embedding for $G$, form from it the dual curve arrangement ${\cal A}(G)$, and use the indices of the curves to assign coordinates to vertices as above. The coordinates of the vertices in a face $f$ may be assigned by referring only to the labels of edges in $f$, in time $O(|f|)$; therefore, all coordinates of $G$ may be assigned in linear time.
The area bound follows easily.
\end{proof}

\begin{figure}[t]
\centering\includegraphics[width=2in]{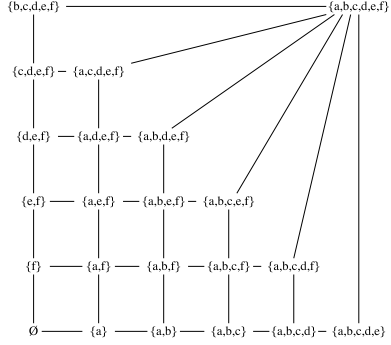}
\caption{The family of sets formed by the union of a prefix and a suffix of some ordered universe forms an $st$-planar learning space with $1+(|U|+1)|U|/2$ states.}
\label{fig:prefixsuffix}
\end{figure}

\begin{corollary}
\label{cor:quadratic}
Any $st$-planar learning space over a set $U$ has at most
$1+(|U|+1)|U|/2$ states.
\end{corollary}

\begin{proof}
Our drawing technique assigns each vertex (other than the topmost one) a pair of coordinates
associated with a pair of elements $\{x_i,y_j\}\subset U$ (possibly with $x_i=y_j$), and each pair of elements can supply
the coordinates for only one vertex. Thus, there can only be one more vertex than subsets of one or two members of $U$.
\end{proof}

The bound of Corollary~\ref{cor:quadratic} is tight, as the family of sets $\cal F$ that are the unions of a prefix and a suffix of a totally ordered set $U$ (Figure~\ref{fig:prefixsuffix}) forms an $st$-planar learning space with exactly $1+(|U|+1)|U|/2$ states.

\section{From Drawings to Quadrant Arrangements}

\begin{figure}[t]
\centering\includegraphics[width=3in]{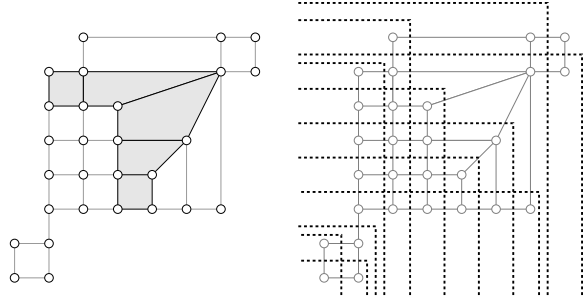}
\caption{Left: A zone in an upright-quad drawing. Right: an arrangement of quadrants through each zone.}
\label{fig:uq2wpla}
\end{figure}

Define a {\em zone} of an upright-quad drawing to be a maximal sequence of interior faces adjacent on opposite sides of each quadrilateral (Figure~\ref{fig:uq2wpla}(left)). A zone consists of a vertical sequence of quadrilaterals sharing horizontal sides and a horizontal sequence of quadrilaterals sharing vertical sides, connected across a diagonal edge; either or both sequence may be empty. We consider any bridge of the graph to form a zone of its own.

\begin{theorem}
\label{thm:uqd2aq}
Each upright-quad drawing is the region graph for an arrangement of quadrants.
\end{theorem}

\begin{proof}
For each zone $z_i$, we choose a coordinate value $x_i$, larger than the $x$-coordinate of the left endpoint of the bottom edge of the zone and (if the bottom edge is non-vertical) smaller than the $x$-coordinate of the right endpoint of the edge.
We similarly choose a coordinate value $y_i$, larger than the $y$-coordinate of the bottom endpoint of the left edge of the zone and (if the left edge is non-horizontal) smaller than the $y$-coordinate of the top endpoint of the edge. We choose these coordinates in such a way that, if zone $i$ meets the right exterior path of the drawing prior to zone $j$, then $x_i<x_j$, and, if zone $i$ meets the left exterior path of the drawing prior to zone $j$, then $y_i<y_j$.  We then draw a curve for zone $z_i$ by combining a horizontal ray left from $(x_i,y_i)$ with a vertical ray down from $(x_i,y_i)$ (Figure~\ref{fig:uq2wpla}(right)). This curve is easily seen to cross all faces of zone $z_i$, and no other interior faces of the drawing; thus, the arrangement $\cal A$ of these curves forms a planar dual to the drawing (except, as before, that it does not have a vertex representing the external face).
\end{proof}

\begin{corollary}
Each upright-quad drawing represents an $st$-planar learning space.
\end{corollary}

\begin{proof}
This follows from Theorem~\ref{thm:qa2stpls} and Theorem~\ref{thm:uqd2aq}.
\end{proof}

As different sets of translation vectors for quadrants form combinatorially equivalent arrangements if and only if the sorted orders of their $y$-coordinates form the same permutations with respect to the sorted order of their $x$-coordinates, a bound on the number of $st$-planar learning spaces follows.

\begin{corollary}
There are at most $n!$ combinatorially distinct $st$-planar learning spaces over a set of $n$ unlabeled items.
\end{corollary}

More precisely, with high probability any permutation corresponds to a learning space with only two combinatorially distinct upright-quad drawings (one formed by flipping the other diagonally), so the number of distinct $st$-planar learning spaces is $\frac12 n! (1-o(1))$.

\section{Conclusions}

We have characterized $st$-planar learning spaces, both in terms of the existence of an upright-quad drawing and as the region graphs of quadrant arrangements.
Our technique for drawing these graphs provides good vertex separation and small area, and it is straightforward to verify from its drawing that a graph is an $st$-planar learning space.

Our results can be viewed as showing that the {\em convex dimension} of an antimatroid is two if and only if its order dimension is two. It is known that the order dimension is always upper bounded by the convex dimension~\cite[Corollary III.6.10]{KorLovSch-91}. However, these two quantities are not always equal. For instance, the antimatroid over $\{0,1,2,3,4\}$ formed by the subsets that either don't contain $0$ or do contain three or more items has order dimension at most five, while it has convex dimension six. We note that the convex dimension is computable in polynomial time as the width of a related poset~\cite[Theorem III.6.9]{KorLovSch-91}, and are hopeful that this result may also lead to interesting methods of graph drawing, analogously to how our minimum-dimensional lattice embedding technique~\cite{Epp-EJC-05} led to drawing algorithms for arbitrary nonplanar partial cubes~\cite{Epp-GD-04}.

Alternatively, when confronted with the task of drawing large nonplanar learning spaces, it may be helpful to find large $st$-planar subgraphs and apply the techniques described here to those subgraphs. Additionally, it would be of interest to extend our $st$-planar drawing techniques to broader classes of partial cubes.

\raggedright
\bibliographystyle{abuser}
\bibliography{media}

\end{document}